\documentclass[12pt]{article}

\title{Some Consequences of Dark Energy Density varying Exponentially with Scale Factor}
\author{Sanil Unnikrishnan$^\dagger$  and T. R. Seshadri $\ddagger$ \\
{}\\
{\it Department of Physics \& Astrophysics,}\\ 
{\it University of Delhi, Delhi 110007,India.}\\
$\dagger${\small e-mail : sanil@physics.du.ac.in}   \\
$\ddagger${\small e-mail :~~ trs@physics.du.ac.in}
}

\usepackage{epsfig}
\begin{document}

\maketitle
\begin{abstract}
In this paper we have explored the consequences of a model of dark energy with
its  
energy density varying 
exponentially with the scale factor. We first consider the model with $ \rho_{\phi} 
\propto e^{\kappa a} $ , where $\kappa $ is a constant. This is a kind of generalisation 
of the cosmological constant model with $\kappa = 0$. 
We show that such an exponentially varying dark energy density with the scale factor 
naturally leads to an equivalent phantom field. We also consider a model with 
$ \rho_{\phi} \propto e^{\kappa /a} $ and we show that this also naturally leads to 
an equivalent phantom field.  
\end{abstract}

\section{Introduction}
Observations  on Supernova Type Ia, CMB and studies on large scale structure 
formation have strongly suggested that at the present epoch the universe is in a phase of 
accelerated expansion \cite{1,2}. Within the standard framework of General Relativity, 
this kind of expansion is not possible with the conventional kind of source, 
{\it i.e.}, which has an equation of state of the form $p=w\rho$ with 
$0 \le w \le 1$.  The present accelerated 
expansion of the universe 
seems to indicate that there exists some form of 
matter-energy with negative pressure which is causing this acceleration. 
(One could, of course, argue that Einstein's general 
relativity may not be the fully correct/complete theory on the cosmological scales \cite{3}.
But we will not consider this possibility in this paper.)
The former possibility has motivated the exploration of alternative kinds 
of sources with $ p < -\frac{1}{3}\rho $ as well as a resurgence of 
investigations using cosmological constant ($w=-1$). Such a matter energy 
component with equation of state $ p < -\frac{1}{3}\rho $ is 
generally known as 'Dark energy'. Several dark energy models have been 
considered in literature. The cosmological 
constant is just the simplest case with 
$w=-1$.  Models with cosmological constant predict a transition to an 
accelerated expansion phase of the Universe. 

Introducing the cosmological constant, however, leads to two serious problems. The 
first one is that the observations indicate that the value of  $ \Lambda $ is 
many orders of magnitude smaller than what is expected from the energy scales 
involved in the Early Universe. This problem is known as the cosmological 
constant problem \cite{4,5,6} . The second problem is that the observations indicate 
that at 
the present epoch the energy density  of the $ \Lambda $ term, {\em i.e.},
$ \rho_{\Lambda} = \frac{\Lambda}{8\pi G}$ is of the same order in magnitude 
as the energy density of matter $\rho_{m}$ with  $ \Lambda $ having begun 
to dominate in the recent past. This requires extreme fine tuning 
of the value of $ \Lambda $. This problem is known as the cosmic coincidence 
problem. The vacuum energy density $\rho_{\Lambda}$ does not change with the 
expansion of the Universe . Hence, there is a need to explore more
general models of dark energy. 

The other models of dark energy in which the density of dark energy,
$\rho_{de}$, changes as the 
universe expands are based on Quintessence, Phantom field, Tachyon field etc. 
Quintessence models refer to those models in which dark energy is represented 
by a standard scalar field with a potential $ V(\phi)$ \cite{7,8,9}. Specific forms of potential 
which satisfy certain conditions  admit tracking solution which may 
possibly solve the cosmic coincidence problem \cite{9b,10}. Quintessence
models have an  
equation of state parameter $w_{de}$ greater than minus one. 
Observational constraints on  $w_{de}$ do not exclude the possibility of 
$w_{de}$ being less than minus one at the present epoch \cite{11,11a}. This leads to the 
possibility that the present cosmic acceleration may be driven by a phantom 
field ($w_{de} < -1$). Energy density of the phantom field increases as the 
universe expands. The phantom energy could be obtained from either a scalar 
field with negative kinetic energy \cite{12,13} term or by some form of non minimal 
coupling of the scalar field with gravity \cite{14}. The phantom field described in this 
paper refer to that with a negative kinetic energy term. 
Some models of phantom energy with a 
constant $w_{de}$ lead to a singularity after a finite time in the future \cite{15,16}. 
This is called the Big Rip.
Certain other models with dynamical $w_{de}$ do not lead to Big Rip \cite{17}.
There is yet another theoretical possibility that the present accelerated 
expansion might be due to tachyon fields \cite{18,19}. These were originally motivated 
from the string theory. In these models $w_{de} > -1$.

In all these scalar field models the equation of state is given as 
$ p_{\phi} = w_{\phi}\rho_{\phi}$. Dark energy with different equation of 
state has also been considered in literature, for example, Chaplygin gas with 
$ p = \frac{\alpha}{\rho}$ \cite{21,22} and models of dark energy with generalised 
equation of state 
$p = \alpha(\rho - \rho_{o}$) \cite{20}.

It has been argued that probing the dark energy via its density evolution 
rather than its equation of state has a number of advantages {\cite{freese}}.
 In this paper we consider a model of dark energy with its energy density 
varying exponentially with the scale factor. After discussing the basic
formalism for a general scalar field in section \ref{emt},
we will consider in section \ref{kappaa}
the model with $ \rho_{\phi} \propto e^{\kappa a}$ where $\kappa$ is a
constant. This is 
a kind of generalisation of the cosmological constant model which corresponds to $\kappa = 0$. 
We will show that such a model naturally 
leads to an equivalent phantom field. This model, however, leads to Big Rip. 
This motivated us to consider another model with 
$ \rho_{\phi} \propto e^{\kappa /a}$ which is discussed in the section \ref{kappabya}. This also 
naturally leads to an equivalent phantom field but it doesn't lead to 
Big Rip. It asymptotically behaves as a de Sitter Universe. The phase space analysis has been
discussed in section \ref{sa}

\section{Energy-momentum Tensor and the Equations of Motion for a general 
scalar field}  \label{emt}

Let  the Lagrangian density of the scalar field be given by ${\cal L}_{\phi}$ and 
that for the rest of the matter by, ${\cal L}_{source}$.
The complete Einstein-Hilbert action for this system 
is then  given by,
\begin{equation}
\mathcal{S} = \int \left[ \frac{-1}{16\pi G}R  + {\cal L}_{\phi} + \mathcal{L}_
{source}\right] \sqrt{-g}~d^{4}x  
\end{equation} 
We employ the metric with signature ($+---$).
The Lagrangian density ${\cal L}_{\phi}$ for a normal scalar field is given by,
\begin{equation}
{\cal L}_{\phi}=\frac{1}{2}\partial_{\mu}\phi\partial^{\mu}\phi -V(\phi)
\end{equation}
The kinetic energy term for this scalar field is positive. Another class of 
scalar fields have been considered in literature for which the kinetic energy 
term is a negative. These are called phantom fields and their Lagrangian 
density is given by,
\begin{equation}
\mathcal{L_{\phi}} = -\frac{1}{2}\partial_{\mu}\phi\partial^{\mu}\phi -V(\phi)
\end{equation}
In order to incorporate both these kinds of scalar fields, we consider an action 
of the form,
\begin{equation}
\mathcal{S} = \int \{ \frac{-1}{16\pi G}R + [ \frac{\alpha}{2}\partial_{\mu}\phi
\partial^{\mu}\phi -V(\phi)]  +  \mathcal{L}_{source}\} \sqrt{-g}d^{4}x  \label{action}
\end{equation}
where $ \alpha $ takes the value $+1$ for normal scalar field (quintessence) 
and  $-1$ for phantom field.    
The energy momentum tensor of the scalar field obtained from the action in 
equation (\ref{action}) is given by:
\begin{equation}
T_{(\phi)}^{\mu \nu} = \alpha\partial^{\mu}\phi\partial^{\nu}\phi  -g^{\mu \nu}[  
\frac{\alpha}{2}\partial_{\mu}\phi\partial^{\mu}\phi -V(\phi)] \label{tabphi}
\end{equation}

Consider the spatially flat FRW metric: 
\begin{equation}
ds^{2} = dt^{2} - a(t)^{2}[dr^{2} + r^{2}d\theta ^{2} + r^{2}\sin{\theta}^{2}d\phi^{2}]
\label{lineE}
\end{equation}

We further assume that the scalar field is homogeneous. The spatial derivatives of 
such a field vanishes and the field depends only on time \textit{i.e} $ \phi = \phi(t)$. 
Using this form of $\phi$ and the metric corresponding to the line element in equation 
(\ref{lineE}), the energy momentum tensor in equation (\ref{tabphi}) assumes a diagonal 
form given by: 
\begin{equation}
T^{\mu}_{(\phi) \nu} = diag(\rho_{\phi},-p_{\phi},-p_{\phi},-p_{\phi}),
\end{equation}
where, 
\begin{equation} 
\rho_{\phi} = \alpha\frac{\dot{\phi}^2}{2} + V(\phi)    \label{rpq} 
\end{equation} 
and 
\begin{equation} 
p_{\phi} = \alpha \frac{\dot{\phi}^2}{2} - V(\phi) \label{ppq}  
\end{equation}
The equation of state is given by $ p_\phi =w_\phi \rho_\phi $ 
where, 
\begin{equation}    w_\phi = \frac{\alpha \frac{\dot{\phi}^2}{2} - V(\phi)}
{\alpha \frac{\dot{\phi}^2}{2} + V(\phi)} \label{wpq} \end{equation}
The dynamics of the Universe is determined by the Friedmann equations:
\begin{eqnarray}
H^{2} &=& \frac{8\pi G}{3}\rho \label{H} \\
\frac{\ddot{a}}{a} &=& -\frac{4\pi G}{3}(\rho + 3p)
\end{eqnarray}
where $\rho$ and $p$ are the total energy density and the total pressure.
( We have assumed a $ k = 0$ universe).  
The field equation for $\phi$ is obtained from the conservation equation 
$T^{\mu \nu}_{\phi \phantom{\mu \nu}{;\nu}} = 0$ and is given by: 
 \begin{equation}
\ddot{\phi} + 3H\dot{\phi} +  \alpha^{-1} \frac{dV}{d\phi} = 0
\end{equation}
Let $\rho  = \rho_m + \rho_{\phi}$  (We have neglected the contribution from 
radiation since we are interested only  in the late time evolution of the 
Universe).  
The scale factor is normalised so as to make $a(t_0) = 1$. 
The density of matter is hence given by
 \begin{equation}  
\rho_m = \rho_{mo}a^{-3} \label{ma}
\end{equation} 
where $ \rho_{mo} $ is the matter density at the present epoch.
 
\section{Model with $ \rho_{\phi}  \propto  e^{\kappa a}$} \label{kappaa}

As we stated in the introduction, our approach in this paper is to start 
with a form for the evolution of the energy density of the scalar field 
rather than the equation of state. This is the approach that has been 
adopted in \cite{freese}. The present paper is in this spirit. We consider a scalar 
field $\phi$ in a potential $V(\phi)$ whose energy 
density varies with scale factor as $\rho_{\phi}  \propto  e^{\kappa a}$. 
Since we have normalised the scale factor to unity today, we may write,
\begin{equation}
\rho_{\phi} = \rho_{\phi 0}e^{\kappa (a - 1)}\label{ra} \label{rhophiexp}
\end{equation} 
where $\rho_{\phi 0}$ is the dark energy density in the present epoch with 
the scale factor today normalised to unity. 
The continuity equation for the scalar field $T^{\mu \nu}_{\phi \phantom
{\mu \nu}{;\nu}} = 0$ implies: 
\begin{equation}
\frac{d\rho_{\phi}}{da}=-\frac{3}{a}(1+w_{\phi})\rho_{\phi}
\end{equation}
where, $w_{\phi}$ is the ratio of the pressure, $p_{\phi}$, to the energy 
density, $\rho_{\phi}$, of the scalar field. Using the form for $\rho_{\phi}$ 
as given in equation (\ref{ra}), we get,
\begin{equation}
 w_{\phi} = -1 - \frac{a\kappa}{3} \label{wa}
\end{equation}

From equation (\ref{wa}) it follows that if $ \kappa < 0 $ then at sufficiently 
large value of the scale factor $ a $ , the ratio of the pressure to density of 
the scalar field,
$w_{\phi}$ would become greater than one. This would violate causality. In 
order to ensure causality we need $w_{\phi} < +1$ and in order to ensure this 
at all future epochs, we require $\kappa$ to be greater than zero.

Using equations (\ref{rpq}) and (\ref{wpq}), we have 
\begin{equation}
\alpha \dot{\phi}^2=(1+w_{\phi}) \rho_{\phi}         \label{alpha_phi_dot}
\end{equation}
Substituting for $w_{\phi}$ from equation(\ref{wa}), we get,
\begin{equation}
\alpha\dot{\phi}^{2} = -\frac{a\kappa}{3}\rho_\phi 
\end{equation}
For $\rho_\phi > 0 $ and $\kappa > 0$ (as required by the causality condition
), $\alpha$ should be negative.
Hence, the kinetic energy term must have a negative sign which fixes $\alpha=-1$. 
Hence such a form of $\rho_\phi$ naturally leads to an equivalent phantom 
scalar field. 

\begin{figure}
\centering{
\scalebox{2}{\rotatebox{270}{
\psfig{file = 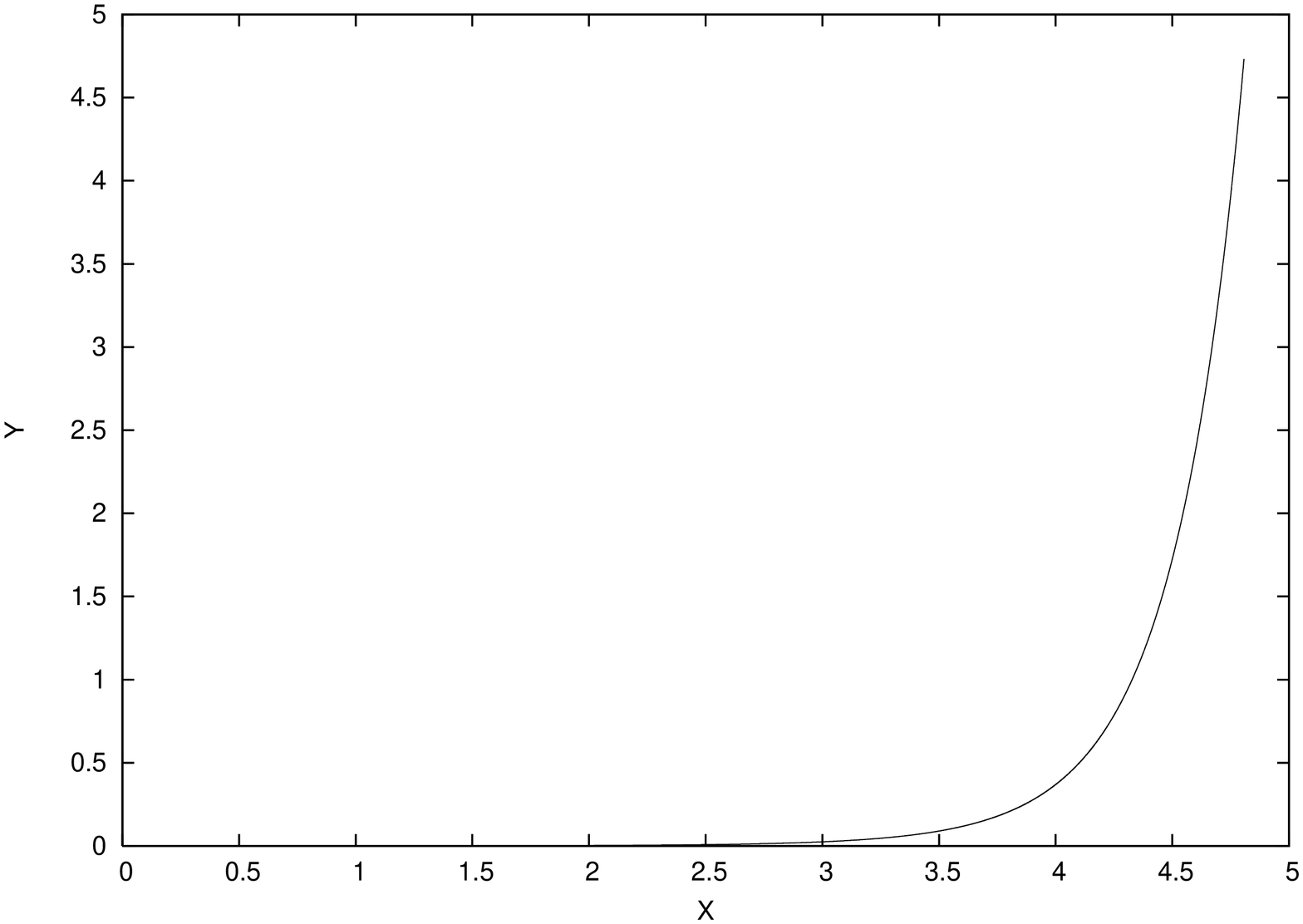, width = 1in}
}}}
\caption{This figure shows the potential $ V(\phi)$ for the 
model $\rho_{\phi} \propto e^{ka}$. In this figure 
$ X = \frac{\phi}{M_{pl}}$ and $Y =
\frac{V(\phi)}{H_{0}^{2}M_{pl}^{2}}10^{-4}$.
For this plot we have taken $\kappa = 1$, $\Omega_{\phi}  = 0.73 $ and  
$\Omega_{m}  = 0.27 $.  }
\label{veka}
\end{figure}

We now explore the nature of the potential $ V(\phi) $ of this phantom field 
which would lead to such a form of $\rho_\phi$. 
From equations (\ref{rpq}), (\ref{ppq}) and (\ref{wpq}) (with $\alpha=-1$) we have,
\begin{eqnarray}
 \rho_{\phi} &=& -\frac{\dot{\phi}^2}{2} + V(\phi) \label{rp} \\
 p_{\phi} &=& -\frac {\dot{\phi}^2}{2} - V(\phi) \label{pp} \\
 w_\phi &=& \frac{- \frac{\dot{\phi}^2}{2} - V(\phi)}{-\frac{\dot{\phi}^2}{2} + 
 V(\phi)} \label{wp} \\
\end{eqnarray}
These equations imply,
\begin{eqnarray}
\dot{\phi}^2 &=& -(1 + w_\phi)\rho_{\phi} \label{phdot} \\
  V(\phi) &=& \frac{(1 - w_\phi)\rho_{\phi}}{2}\label{v}
\end{eqnarray}
The above equation (\ref{v}) together with equations (\ref{rhophiexp}) and
(\ref{wa}) gives the potential $V(\phi)$ as a function of the scale factor ${a}$. 
The derivative of $\phi$ with respect to  scale factor $a$ is given by:
\begin{equation}
\frac {d\phi}{da} = \frac{\dot{\phi}}{\dot a} =  \frac{\dot{\phi}}{aH}\label{dphia}
\end{equation}
where $ H  $ is given  by the Friedmann equation (\ref{H}): 
\begin{equation}
 H^2 = \frac{8\pi G}{3}(\rho_{m0}a^{-3} + \rho_{\phi 0}e^{\kappa(a - 1)})\label{Ha}
\end{equation}
 Using equations (\ref{ra}), (\ref{wa}), (\ref{phdot}), (\ref{v}), (\ref{dphia}) 
 and  (\ref{Ha}) one obtains,
\begin{eqnarray}
\frac{d}{da}(\frac{\phi}{M_{p}}) &=& \frac{ a(\kappa \Omega _{\phi})^{1/2}}{(\Omega _
{\phi}a^3 + \Omega _{m}e^{-\kappa(a - 1)})^{1/2}}\label{pha} \\
V(a) &=& 3\Omega _{\phi}H_{0}^{2}M_{p}^{2}(1 + \frac{a\kappa}{6})e^{\kappa(a - 1)}. \label{va} 
\end{eqnarray}
Here,
\begin{equation} \Omega _{\phi} = \frac{\rho_{\phi_0}}{3 H_{0}^{2}M_{p}^{2}} 
\nonumber 
\end{equation} 
and 
\begin{equation} \Omega _{m} = \frac{\rho_{m_0}}{3 H_{0}^{2}M_{p}^{2}} 
\nonumber 
\end{equation}  
And  
\begin{equation} 
M_{p}^2 = \frac{1}{8\pi G} \nonumber 
\end{equation}
which is the reduced Planck mass in units of $\hbar = c = 1$ . $ H_0 $ 
is the Hubble parameter. 
Using equations (\ref{pha}) \& (\ref{va}) one can plot potential $ V(\phi)$ as 
a function of $\phi$. This plot is shown in figure (\ref{veka}). 

In this plot we have chosen initial condition such that $ \phi = 0$ at $ a =
10^{-5}$ . The field $ \phi $ grows from  $ \phi = 0$ to a value $ \phi =
0.55M_{p}$ at the present epoch ($ a = 1$).

At low value of the field \textit{i.e.} 
for $\phi \ll M_{p}$ which occurs at $ a \ll 1$,
  the differential equation (\ref{pha}) can be approximated as
\begin{equation}
\frac{d}{da}(\frac{\phi}{M_{p}})   \simeq   a\sqrt{\frac{\kappa \Omega
    _{\phi}}{\Omega _{m}e^{\kappa}}} \label{lowphi},
\end{equation}
which on integration gives: 
\begin{equation}
a = \sqrt{\frac{\alpha}{M_{p}}(\phi -\phi_{0})} \label{lowphia}
\end{equation}
Where $\phi_{0}$ is the constant of integration and
\begin{equation}
\alpha = 2\sqrt{\frac{\Omega _{m}e^{\kappa}}{\kappa \Omega
    _{\phi}}}
\end{equation}
 On substituting equation (\ref{lowphia}) in the equation (\ref{va}), we
obtain an approximate functional form of $ V(\phi) $ for small value of the field ($\phi \ll
M_{p}$). This is given as:

\begin{equation}
V(\phi) = V_{0}(1 + \frac{\kappa}{6} \sqrt{\frac{\alpha}{M_{p}}(\phi
  -\phi_{0})})exp(\kappa \sqrt{\frac{\alpha}{M_{p}}(\phi -\phi_{0})}),
\end{equation}
where
\begin{equation}
V_{0} = 3\Omega _{\phi}H_{0}^{2}M_{p}^{2}e^{-\kappa}
\end{equation}
   
Similarly for large value of the field, \textit{i.e.} for $\phi >
M_{p}$, which occurs at $a > 1 $ (and hence at a future epoch), the differential equation (\ref{pha}) can be approximated as: 

\begin{equation}
\frac{d}{da}(\frac{\phi}{M_{p}})   \simeq   \sqrt{\frac{\kappa}{a}} \label{highphi},
\end{equation}
which on integration gives:
\begin{equation}
a = \frac{(\phi -\phi_{0})^{2}}{4\kappa M_{p}^2} \label{highphia}.
\end{equation}
Here $\phi_{0}$ is the constant of integration. Substituting
equation (\ref{highphia}) in the equation (\ref{va}), we obtain an approximate
functional form of the potential  $V(\phi)$ at large value of the field
$\phi$. This is given as:
\begin{equation}
V(\phi) = V_{0}( 1 + \frac{(\phi -\phi_{0})^{2}}{24M_{p}^2})exp(\frac{(\phi
  -\phi_{0})^{2}}{4M_{p}^2}) \label{vahigh}
\end{equation}
We will consider this form of potential given by equation (\ref{vahigh}) in
section \ref{sa} to study the phase space analysis.
 
The value of the constant $\kappa$ can be determined from the epoch of equality of 
matter and dark energy density. Let $z_{m\phi}$ be the red shift of the epoch of 
matter-dark energy equality. Equating the expression for $\rho_{\phi}$ and 
$\rho_{m}$ from equations (\ref{ra}) and (\ref{ma}) we get,
\begin{equation}
\kappa = -(1 + \frac{1}{z_{m\phi}})\mathit{ln}(\frac{\Omega_m}{\Omega_{\phi}}(1 + 
z_{m\phi})^3)\label{k} 
\end{equation}
As $\kappa > 0$ equation(\ref{k}) implies, 
\begin{equation}
z_{m\phi} < (\frac{\Omega_{\phi}}{\Omega_m})^{1/3} - 1\label{z}
\end{equation}
As $\Omega_{\phi} = 0.73$ and $\Omega_{m} = 0.27$ (as indicated by the observations by WMAP \cite{WMAP})
we have $z_{m\phi} < 0.393$. 
This model (with $\rho_{\phi} \propto e^{\kappa a}$), thus, constrains 
the value of $z_{m\phi}$. 
We can see from the expression of $w_{\phi}(a)$ given by equation(\ref{wa}) 
that $ a \rightarrow \infty \Rightarrow w_{\phi}(a) \rightarrow -\infty $.
From the expression of $\rho_{\phi}$ (\ref{ra}) and the equation of state we 
see that $ \rho_{\phi} \rightarrow \infty $ and $p_{\phi}  \rightarrow 
-\infty $ as $ a \rightarrow \infty $. This would then mean that the trace of 
the energy momentum tensor $T=T^{\mu}_{\mu} = \rho_{\phi} - 3p_{\phi}$ 
(and hence the Ricci scalar $R \propto T$) would blow up as 
$ a \rightarrow \infty $. It is of interest to ascertain the epoch when the scale factor 
tends to infinity. In particular, we would like to know whether this happens at a finite 
time or
asymptotically in the infinite future. 

The epoch, $t_{\infty}$, when the scale factor becomes infinitely large can be calculated from the Friedman equation (\ref{Ha}) to be,
\begin{equation}
t_{\infty} = t_{0} + \frac{1}{H_{0}}\int_{1}^{\infty}\frac{da}{a(\sqrt{\Omega_{m}a^{-3} 
+ 
\Omega_{\phi}e^{\kappa(a - 1)}})}\label{t}
\end{equation}
where $t_{\infty}$ is the epoch when $a \longrightarrow \infty$. As the scale factor
becomes infinitely large, the first term under the square root term becomes subdominant 
as compared to the second and so the integral remains finite.
Thus this model exhibits a singularity at a finite future epoch. This is popularly 
known as Big Rip and our model $\rho_{\phi} \propto e^{\kappa a}$ exhibits
that feature.

\section{Model with $ \rho_{\phi}  \propto  e^{\kappa /a}$}  \label{kappabya}

\begin{figure}
\centering
\scalebox{2}{
\rotatebox{270}{
\psfig{file = 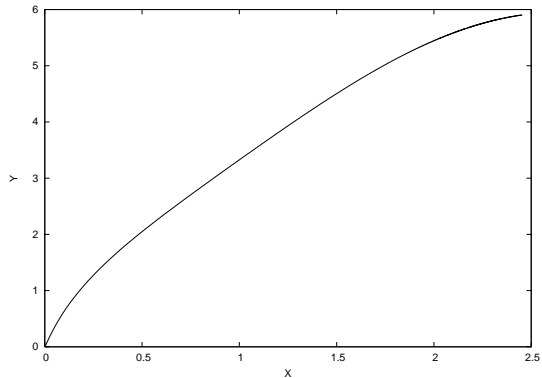, width = 1in}
}}
\caption{This figure shows the potential $ V(\phi)$ for the model 
$\rho_{\phi} \propto e^{k/a}$. In this figure $ X = \frac{\phi}{M_{pl}}$ 
and $Y = \frac{V(\phi)}{H_{0}^{2}M_{pl}^{2}}$. For this plot we have taken 
$\kappa =- 1$, $\Omega_{\phi}  = 0.73 $ and  $\Omega_{m}  = 0.27 $.}
\label{vekbya}
\end{figure}

Let $\rho  = \rho_m + \rho_{\phi}$ where $ \rho_{\phi}  \propto  e^{\kappa /a}$.
Denoting by $\rho_{\phi 0}$ the density of the scalar field in the present
epoch corresponding to $a = 1$, we have:
\begin{equation}
 \rho_{\phi} = \rho_{\phi 0}e^{{\kappa}\frac{1 - a}{a}}\label{ra1} 
\end{equation}
We again consider the action of the form (\ref{action}).
The continuity equation $T^{\mu \nu}_{\phi \phantom{\mu \nu}{;\nu}} = 0$ 
would imply that: 
\begin{equation}
 w_{\phi} = - 1 + \frac{\kappa}{3a} \label{wa1}
\end{equation}
If $ \kappa > 0 $ then at sufficiently low value of the scale factor $ a $, $w_{\phi}$ 
would become greater than one. This would violate causality. So for $w_{\phi}$ to 
be less than one requires $\kappa$ to be less than zero. 
Hence in this model causality requires that $ \kappa < 0 $. With these constraints,  
we investigate the sign of the kinetic energy term of the 
scalar field that would lead to $ \rho_{\phi}$ of the form (\ref{ra1}) and also
calculate the form of the potential $V(\phi)$ of the field $\phi$. 
Using equations (\ref{rpq}), (\ref{ppq}), (\ref{wpq}), (\ref{ra1}) \& (\ref{wa1}) we 
obtain,
\begin{equation}
\alpha \dot{\phi}^{2} = \frac{\kappa}{3a}\rho_\phi 
\end{equation}
For $\rho_\phi > 0 $ and $\kappa < 0$ (we have shown that for this model causality 
requires $\kappa < 0$),
the left hand side of the equation is negative. So in this case too, the kinetic energy term must be negative. 
Hence such a form of $\rho_\phi$ also naturally leads to an equivalent phantom scalar 
field. 
We now obtain the form of the potential $ V(\phi) $ of the phantom field which would 
lead to such a form of $\rho_\phi$. 
For  $\rho_{\phi}$ given by equation (\ref{ra1}), the Friedmann equation (\ref{Ha}) 
becomes: 
\begin{equation}
 H^2 = \frac{8\pi G}{3}(\rho_{m0}a^{-3} + \rho_{\phi 0}e^{\kappa\frac{1 - a}{a}})
 \label{Ha1}
\end{equation}
 Using equations (\ref{ra1}), (\ref{wa1}), (\ref{phdot}), (\ref{v}),
 (\ref{dphia}) and (\ref{Ha1}) 
 we obtain, 
\begin{eqnarray}
\frac{d}{da}(\frac{\phi}{M_{p}}) &=&  \sqrt{\frac{ -\kappa \Omega _{\phi}}{\Omega _
{\phi}a^3 + \Omega _{m}e^{-\kappa(\frac{1}{a} - 1)}}\label{pha1}} \\
\frac{V(a)}{H_{0}^{2}M_{p}^{2}} &=& 3\Omega _{\phi}(1 - \frac{\kappa}{6a})e^{\kappa(\frac{1}{a} - 
1)}\label{va1} 
\end{eqnarray}
Using equations (\ref{pha1}) \& (\ref{va1}) one can plot potential $ V(\phi)$ as a 
function of $\phi$. This plot is shown in figure (\ref{vekbya}).
In this plot also we have chosen an initial condition such that $ \phi = 0$
at $ a = 10^{-5}$ . This gives a value of the field $ \phi = 0.7M_{p}$ at the
present epoch ($ a = 1$). The field $ \phi $ grows from  $ \phi = 0$ to a
value $\phi \simeq 2.45M_{p}$. As done in the
previous section we obtain the limiting form of the potential $ V(\phi)$ from
equations (\ref{pha1}) and (\ref{va1}).

When $ \phi \ll M_{p}$ which occurs at $ a \ll 1$, the potential $ V(\phi)$ is nearly flat.
When  $ \phi > M_{p}$ and $ a \gg 1$, the equation (\ref{pha1}) can be
approximated as:
\begin{equation}
\frac{d}{da}(\frac{\phi}{M_{p}})   \simeq   \sqrt{ \frac{-\kappa}{a^{3}}} \label{highphi1}, 
\end{equation}
which on integration gives: 
\begin{equation}
a = \frac{-4\kappa M_{p}^{2}}{(\phi -\phi_{0})^{2}} \label{highphia1}.
\end{equation}
Here $ \phi_{0}$ is the constant of integration. Substituting this equation
in equation (\ref{va1}), we obtain a limiting form of the potential at large
value of the scale factor and for $ \phi > M_{p}$. This is given as:
\begin{equation}
V(\phi) = V_{0}( 1 + \frac{(\phi -\phi_{0})^{2}}{24M_{p}^2})exp(-\frac{(\phi
  -\phi_{0})^{2}}{4M_{p}^2}) \label{vahigh1}
\end{equation} 
where 
\begin{equation}
V_{0} = 3\Omega _{\phi}H_{0}^{2}M_{p}^{2}e^{-\kappa}
\end{equation}
It is interesting to note that this limiting form of the potential given by
equation (\ref{vahigh1})  is very similar to that obtained in the previous
section ( equation (\ref{vahigh})) with the only difference in the sign in the
exponential part. 
We will also consider this form of the potential given by equation
(\ref{vahigh1}) in section \ref{sa} to study the phase space analysis.  

In this model also we can obtain the value of the constant $\kappa$ from the red shift 
$z_{m\phi}$ of the epoch of matter-dark energy equality. Equating the expressions 
for  $\rho_{\phi}$ and $\rho_m$ we obtain: 
 \begin{equation}
\kappa =  \frac{1}{z_{m\phi}}\mathit{ln}(\frac{\Omega_m}{\Omega_{\phi}}(1 + z_{m\phi})^3)
\label{k1} 
\end{equation}
As $\kappa < 0$, the above equation (\ref{k1}) would imply that: 
 \begin{equation}
z_{m\phi} < (\frac{\Omega_{\phi}}{\Omega_m})^{1/3} - 1
\end{equation}
This is same as equation (\ref{z}). Hence we can see that both this model and the model considered in the previous section 
puts same constraints on $z_{m\phi}$. Using equations (\ref{ra1}) 
\& 
(\ref{wa1}) we can see that this model ($\rho_{\phi} \propto e^{\kappa /a}$) 
asymptotically 
behaves as a De Sitter's Universe. So this model does not lead to Big Rip.
This is unlike the previous model which exhibits a Big Rip.
\section{Phase space analysis} \label{sa}

As the Universe expands, the relative contribution of pressure-less matter keeps 
decreasing and as the scale factor becomes very large, the matter content can be 
neglected. Let us consider the model with $\rho_{\phi} \propto e^{\kappa a}$ and analyse the situation in this limit of large $a$. In this limit,
we neglect the energy density contribution of matter and only consider that due to scalar 
field and integrate the equation of motion for the scalar field to obtain. 
\begin{equation}
\phi(a) = 2\sqrt{\frac{\kappa a}{8\pi G}} + \phi_{0}
\end{equation}
where $\phi_{0}$ is the constant of integration. Using this equation and the equation (\ref{va}) we obtain the following equation:
\begin{equation}
V(\phi) = V_{0}(1 + \frac{\phi^2}{24M_p^2})exp(\frac{\phi^2}{4M_p^2})\label{V1}
\end{equation}
where \begin{equation} V_{0} = 3H_{0}^{2}M_{p}^{2}e^{-\kappa} \end{equation}
(Equation (\ref{V1}) is the same as the equation (\ref{vahigh}) but in the equation (\ref{V1}) we have taken $\phi_{0}$ to be zero ).
From equation (\ref{wa}) the redshift dependence of $w_{\phi}(a)$ is of the form, $ w_{\phi} = -1 + ma $ 
where m is a constant and $a$ is the scale factor, together with the continuity equation,
\begin{equation}
\frac{d}{da}(\rho a^3) +3pa^2 = 0 \label{ce}
\end{equation}
this form of $w_{\phi}(a)$ imposes, $ \rho_{\phi} = Ae^{-3ma} $. So in order to
investigate whether different initial condition leads to  $\rho_{\phi}$ of
the form $\rho_{\phi} \propto e^{\kappa a}$, we plot $w_{\phi}$ as a function of the
scale factor for different initial conditions.
\begin{figure}
\centering
\scalebox{2}{
\rotatebox{270}{
\psfig{file = 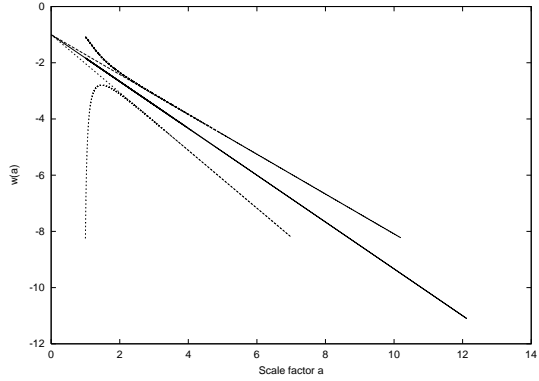, width = 1in}
}}
\caption{Plot of equation of state parameter $w(a)$ verses scale factor for 
different initial conditions for the potential given by equation (\ref{V1}). 
Here $\rho_{m} = 0$ and $\kappa = 2.5$.  }
\label{saek1}
\end{figure}
This plot is shown 
in fig(\ref{saek1}). From this we can conclude that for different 
initial condition $w(a)$ is of the form $ w(a) = -1  +  ma$ asymptotically with the 
constant $m$ depending on the initial condition. Hence, asymptotically $\rho_{\phi}$ 
will be of the form $ \rho_{\phi} = Ae^{\kappa a} $ for different initial conditions but both the 
constants $ \kappa  $  and $ A $ depends on the initial condition. Phase portrait (for 
the case when the potential is given by equation (\ref{V1})) is shown in fig(\ref{saek}). 
\begin{figure}
\centering
\scalebox{2}{
\rotatebox{270}{
\psfig{file = 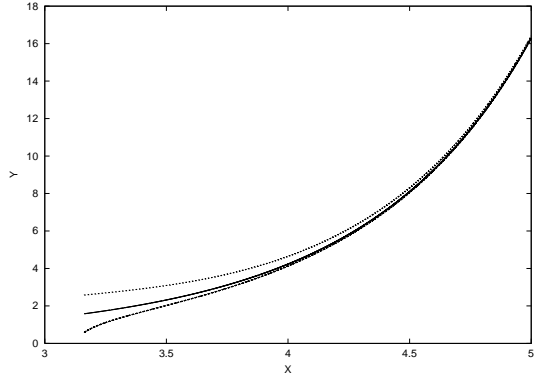, width = 1in}
}}
\caption{Phase portrait in model with potential given by equation (\ref{V1}). 
Here $\rho_{m} = 0$ and $\kappa = 2.5$. The initial condition chosen is the same 
as in fig(\ref{saek1}).  Here $X=\phi/M_p$ and $Y=\dot\phi/{H_0 M_p}$  }
\label{saek}
\end{figure}

\begin{figure}
\centering
\scalebox{2}{
\rotatebox{270}{
\psfig{file = 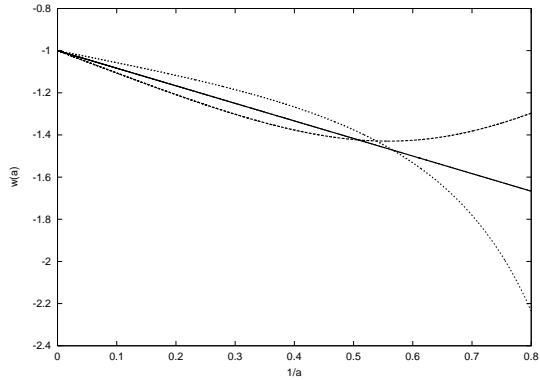, width = 1in}
}}
\caption{Plot of equation of state parameter $w(a)$ verses inverse of scale factor for different 
initial conditions for the potential given by equation (\ref{V2}). Here $\rho_{m} = 0$ 
and $\kappa = -2.5$. Here $1/a \longrightarrow 0 \Rightarrow$ assymptotically into the future. }
\label{saekbya1}
\end{figure}

\begin{figure}
\centering
\scalebox{2}{
\rotatebox{270}{
\psfig{file = 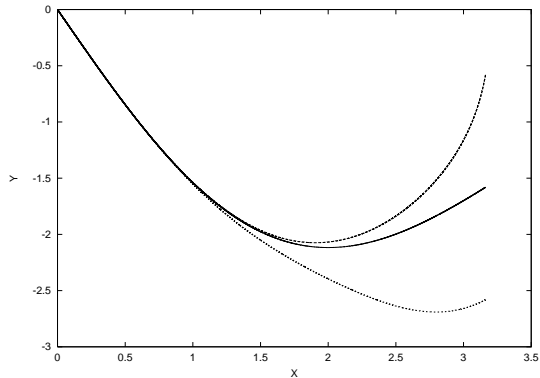, width = 1in}
}}
\caption{Phase portrait in model with potential given by equation (\ref{V2}). 
Here $\rho_{m} = 0$ and $\kappa = -2.5$. The initial condition chosen is the same as in 
fig(\ref{saekbya1}). Here $X=\phi/M_p$ and $Y=\dot\phi/{H_0 M_p}$}
\label{saekbya}
\end{figure}

Similarly for the model $\rho_{\phi} \propto e^{\kappa /a}$, in the limit of large $a$, we get,
 
 \begin{equation}
V(\phi) = V_{0}(1 + \frac{\phi^2}{24M_p^2})exp(-\frac{\phi^2}{4M_p^2})\label{V2}
\end{equation}
where \begin{equation} V_{0} = 3H_{0}^{2}M_{p}^{2}e^{-\kappa} \end{equation}
(Equation (\ref{V2}) is the same as the equation (\ref{vahigh1}) but in the equation (\ref{V2}) we have taken $\phi_{0}$ to be zero ).
Here we plot $w(a)$ verses $1/a$ for different initial conditions. This plot is shown in the fig(\ref{saekbya1}). 
Here also we can 
conclude that for different initial conditions $w(a)$ is of the form $ w(a) = -1 + 
\frac{m}{a}$ asymptotically with the constant $m$ depending on the initial condition. We conclude that asymptotically 
$\rho_{\phi}$ will be of the form $ \rho_{\phi} = Ae^{\kappa /a} $ for different initial 
conditions but both the constants $ \kappa$ and $ A$  depends on the initial condition. 
Phase portrait (for the case when the potential is given by equation (\ref{V2})) is shown in 
fig(\ref{saekbya}).
   
\section{Conclusions}

In this paper we have shown that the dark energy density varying exponentially 
as $\rho_{\phi} \propto e^{\kappa a}$ and $\rho_{\phi} \propto e^{\kappa /a}$ 
naturally lead to an equivalent phantom field. We have shown in section \ref{sa} 
that in the limit of large $a$ phantom potential of the form given by equation (\ref{V1})
leads to $\rho_{\phi} 
\propto e^{\kappa a}$ for a particular initial condition on $\phi$ and $\dot{\phi}$. 
For other nearby initial conditions $\rho_{\phi}$ tends to this form
asymptotically and  the constant $\kappa$ depends on the initial condition ($\phi$ 
and $\dot{\phi}$). This model leads to Big Rip. Similarly in the large $a$ limit, phantom potential of the form given by equation (\ref{V2}) leads to 
$\rho_{\phi} \propto e^{\kappa /a}$ for certain initial conditions. For 
other nearby initial conditions  $\rho_{\phi}$ approaches the form $\rho_{\phi} \propto e^{\kappa /a}$
asymptotically and  the constant $\kappa$ depends on the initial conditions 
($\phi$ and $\dot{\phi}$). It does not leads to Big Rip but rather it 
asymptotically leads to De Sitter Universe. 

\begin{flushleft}
\textbf{ Acknowledgement}
\end{flushleft}
S.U. thanks C.S.I.R, India for a Junior Research Fellowship.
TRS thanks IUCAA for the support provided
through the Associateship Program and the facilities at
the IUCAA Reference Centre at Delhi University


\newpage

\end{document}